\documentclass[conference,10pt]{IEEEtran}

\usepackage{amsmath,graphicx}
\usepackage{amssymb}
\usepackage{lineno,hyperref}
\usepackage{url}
\usepackage{float}
\usepackage{textcomp}
\usepackage{xcolor}
\usepackage{algorithm}
\usepackage{subcaption}
\usepackage{algpseudocode}
\usepackage{lipsum}
\usepackage{multirow}
\usepackage{enumitem}
\usepackage{cite}

\usepackage{titlesec}
\usepackage{lipsum} 
\usepackage{graphicx}
\usepackage{amsmath}
\usepackage{cite}
\usepackage{cuted}

\renewcommand\thesection{\arabic{section}}
\renewcommand\thesubsection{\thesection.\arabic{subsection}}
\renewcommand\thesubsubsection{\thesubsection.\arabic{subsubsection}}

\titleformat{\section}
  {\normalfont\bfseries\fontsize{12}{14}\selectfont}
  {\thesection}{1em}{}
  
\titlespacing*{\section}
  {0pt}{0pt}{0pt}  
  
\titleformat{\subsection}
  {\normalfont\itshape\fontsize{10}{12}\selectfont}
  {\thesubsection}{1em}{}
\titlespacing*{\subsection}
  {0pt}{0pt}{0pt}  

\titleformat{\subsubsection}[runin]
  {\normalfont\itshape\fontsize{10}{12}\selectfont}
  {\thesubsubsection:}{0.5em}{}[.]
\titlespacing*{\subsubsection}
  {0pt}{0pt}{0pt}  

\usepackage[none]{hyphenat}

\begin{document}

\title{\fontsize{18pt}{22pt} \textbf{Adaptive Spatial-temporal Estimation on the Graph Edges via Line Graph Transformation}
    \vspace{-10pt}

}

\author{
    \textbf{\textit{Yi Yan\textsuperscript{1}, Ercan E. Kuruoglu\textsuperscript{1}\textsuperscript{*}}\vspace{10pt}
}
    \\
    \textsuperscript{1}\textit{Institute of Data and Information, Shenzhen International Graduate School, Tsinghua University} \\
    \textit{\textsuperscript{*}Corresponding author: kuruoglu@sz.tsinghua.edu.cn}
}

\maketitle
{\renewcommand\thefootnote{}\footnotetext{Yi Yan was affiliated with Tsinghua University during the completion of this work. This paper is a preprint of a paper submitted to the IET International Radar Conference (IRC 2025) and is subject to Institution of Engineering and Technology Copyright. If accepted, the copy of record will be available at IET Digital Library.}}

\pagestyle{empty}
\thispagestyle{empty}



\begin{strip}
    \vspace{-60pt}
    \begin{center}
        \noindent\textbf{Keywords:} \MakeUppercase{Graph Signal Processing, Time-varying signal, Online estimation, Line graph, Spatial-temporal estimation}
    \end{center}

    \section*{Abstract}
    \vspace{10pt} 
    Spatial-temporal estimation of signals on graph edges is challenging because most conventional Graph Signal Processing techniques are defined on the graph nodes. Leveraging the Line Graph transform, the Line Graph Least Mean Square (LGLMS) algorithm unifies the Line Graph transformation with classical adaptive filters, reinterpreting online estimation techniques for time-varying signals on graph edges. LGLMS leverages the full power of existing GSP techniques on signals on edges by embedding edge signals into node representations, eliminating the necessity of redefining edge-specific techniques. Experimenting with transportation graphs and meteorological graphs, with the signal observations having noisy and missing values, we confirmed that LGLMS is suitable for the online prediction of time-varying edge signals.
    \vspace{3pt} 
\end{strip}

\section{Introduction}
Effectively processing edge signals in graph-structured data remains a fundamental challenge in Graph Signal Processing (GSP).  
While nodes signal representation naturally is well-suited for applications such as social networks \cite{Fan_2023_social}, biological interactions \cite{brain_modeling, zhao2023sequential}, and transportation systems \cite{transportation}, a major challenge lies in effectively processing and interpreting signals that particularly reside on the edges rather than the nodes \cite{yan2024signal}. 
This edge-based signal representation is critical in applications such as modeling population mobility during pandemics \cite{Panagopoulos_2021_covid} or tracking ocean drifter trajectories to understand climate dynamics \cite{Schaub_2020_Random}.
By developing robust methods to process edge-based signals, researchers can unlock new insights into complex systems, paving the way for innovative solutions to pressing global problems, such as improving pandemic response strategies, enhancing climate change resilience,  mitigating traffic congestion, and diagnosing neurological disorders.

The smoothness assumption of graph signals, typically defined on nodes, has played a key role in GSP, enabling tasks like classification, regression, and clustering on graphs \cite{dong2020graph, Leus_2023_GSP}. 
In GSP, smoothness is often interpreted as graph signals to be dominated by lower-frequency components \cite{Ortega_2018}.
However, this assumption can be problematic, particularly in the presence of heterophily, where connected nodes (or edges) may have significantly different signal values (non-smooth) \cite{luan2022revisiting_heterophily, wang_2024_understanding_heterophily}. 
Additionally, smoothness-based GSP techniques are usually defined on the graph nodes; the challenge of the smoothness assumption on graph edges is not straightforward and demands alternative solutions. 
One approach for processing edge signals is the Hodge-Laplacian framework, which represents edges as simplicial complexes and decomposes signals into gradient, curl, and harmonic components \cite{Barbarossa_2020, Schaub_2021_higher_order_networks, Yang_2022_Simplicial, yan2025binarized}. 
Similar operations can be conducted on the cell complex and generalized cell complex as well \cite{Roddenberry_cell_complex_2022, Stefania_2024_cell_complex, Marinucci_2024_cell_complex}.
However, the Hodge Laplacian approach still relies on specific assumptions about edge signals, which may not always align with real applications. 
In particular, it assumes that signals on edges can be meaningfully decomposed in this manner, which might not be generalizable to all applications.

Beyond static edge signals, many real-world applications involve time-varying data on graph edges, such as water flow in rivers \cite{Krishnan_2023_SVAR} and traffic flow on road networks \cite{schaub2018flow}. 
The Simplicial Vector Autoregressive (SVAR) model is proposed to tackle the time-varying task by redefining the VAR model on simplicial complexes using the Hodge Laplacians \cite{Krishnan_2023_SVAR}. 
However, proper deployment of the SVAR model requires a complicated process of learning model parameters from the data, not to mention that the Hodge Laplacian methods require a redefinition of edge operations compared to GSP methods. 
A more fundamental alternative is to rethink GSP itself for edge-based learning, moving beyond node-centric models to develop a framework that inherently captures the structure and dynamics of time-varying edge signals.

We propose the Line Graph Least Mean Squares (LGLMS) algorithm for the online estimation of time-varying signals on graph edges under the scenario where noise and missing observations corrupt the signal. 
LGLMS embeds edge signals as node signals using the Line Graph transformation and applies a bandlimited filter to process them. 
LGLMS is the first framework to unify well-established GSP concepts with classical adaptive filtering concepts into a cohesive algorithm for real-time edge signal estimation, enabling direct signal processing on graph edges rather than only graph nodes.
The LGLMS algorithm is tested and confirmed to accurately predict noisy time-varying traffic data and meteorological data on the graph edges under the various missing observation scenarios modeled by smoothness and random edge sampling.

\section{Background and Notation}
\label{sec_Background}
A graph $\mathcal{G}$ is formed by $N_n$ nodes and $N_e$ edges. 
In this paper, the graphs are assumed to be unweighted and undirected.  
The subscript $n$ denotes nodes and $e$ denotes edges.
The node-to-edge incidence is recorded in the incidence matrix $\mathbf{B}\in \mathbb{R}^{N_n \times N_e}$. 
The rows of $\mathbf{B}$ are associated with the nodes, and the columns of $\mathbf{B}$ are associated with the edges. 
If a node $v_i$ is connected to an edge $e_j$, then the associated $ij^{th}$ entry in $\mathbf{B}$ will have a magnitude of 1. 
One of the most essential operations in GSP is the Graph Fourier transform (GFT), which is defined on the eigendecomposition of the graph Laplacian matrix $\mathbf{L}\in \mathbb{R}^{N_n \times N_n}$. 
$\mathbf{L}$ is the difference between the degree matrix $\mathbf{D}$ and the adjacency matrix $\mathbf{A}$.
By definition, the $ij^{th}$ element of $\mathbf{A}$ is $1$ when there is an edge between node $v_i$ and node $v_j$.
The degree matrix $\mathbf{D}$ is a diagonal matrix that is formed by recording the diagonal entries as the sum of all elements along the rows of $\mathbf{A}$. 
A random orientation is assigned (this is not the edge direction) to the edges in $\mathbf{B}$, so if an edge goes from node $i$ to node $j$, the corresponding entry in $\mathbf{B}$ will be 1 and if the edge is leaving relative to the orientation the value will be -1 \cite{Schaub_2020_Random, schaub2018flow}. 
This leads to the equality $\mathbf{L} = \mathbf{BB}^T$.

\begin{figure}[h]
     \centering
     \begin{subfigure}{0.5\textwidth}
         \centering
         \includegraphics[width=\textwidth]{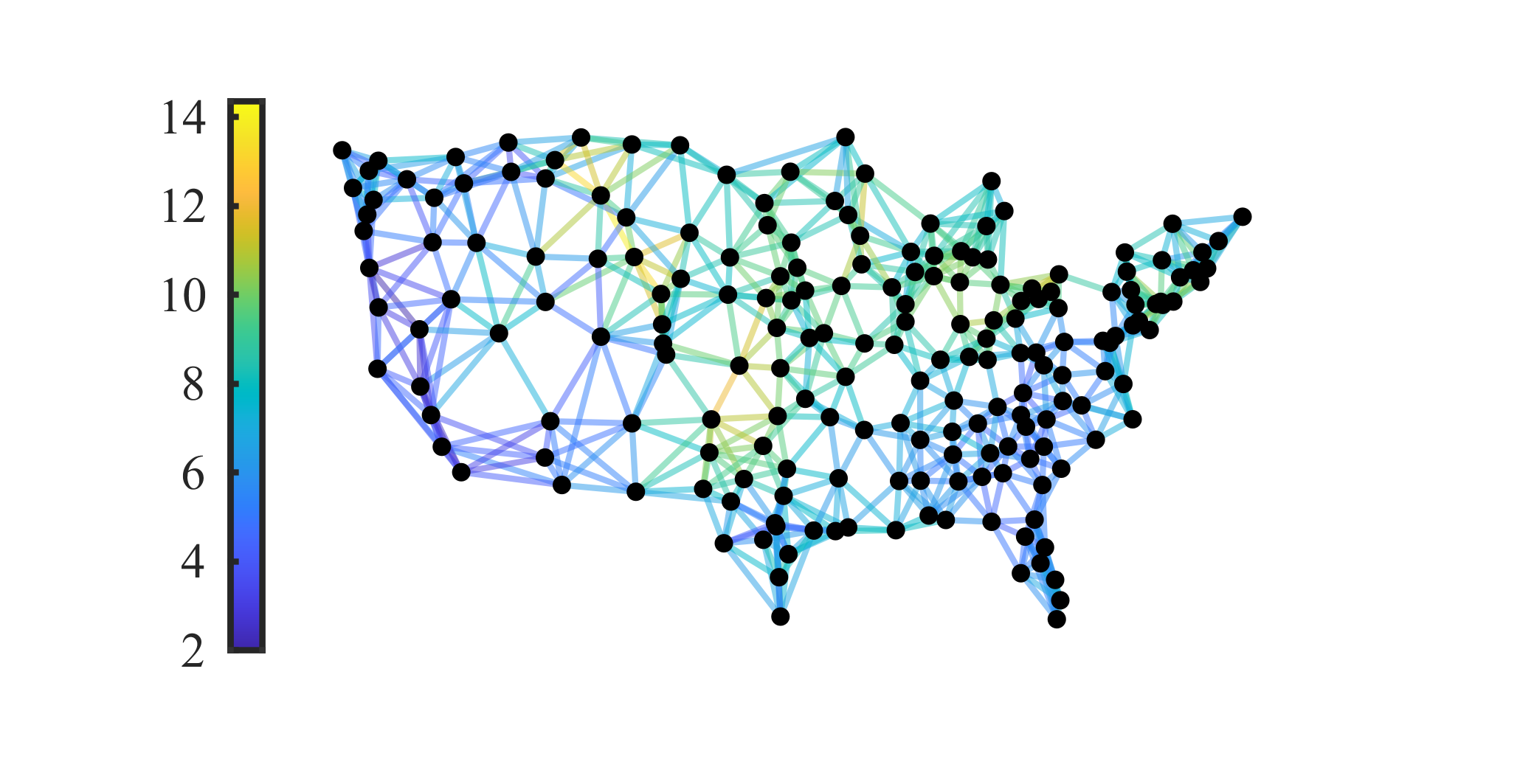}
                  \vspace{-30pt}
         \caption*{$t=1$}    
     \end{subfigure}      
     \begin{subfigure}{0.5\textwidth}
         \centering
         \includegraphics[width=\textwidth]{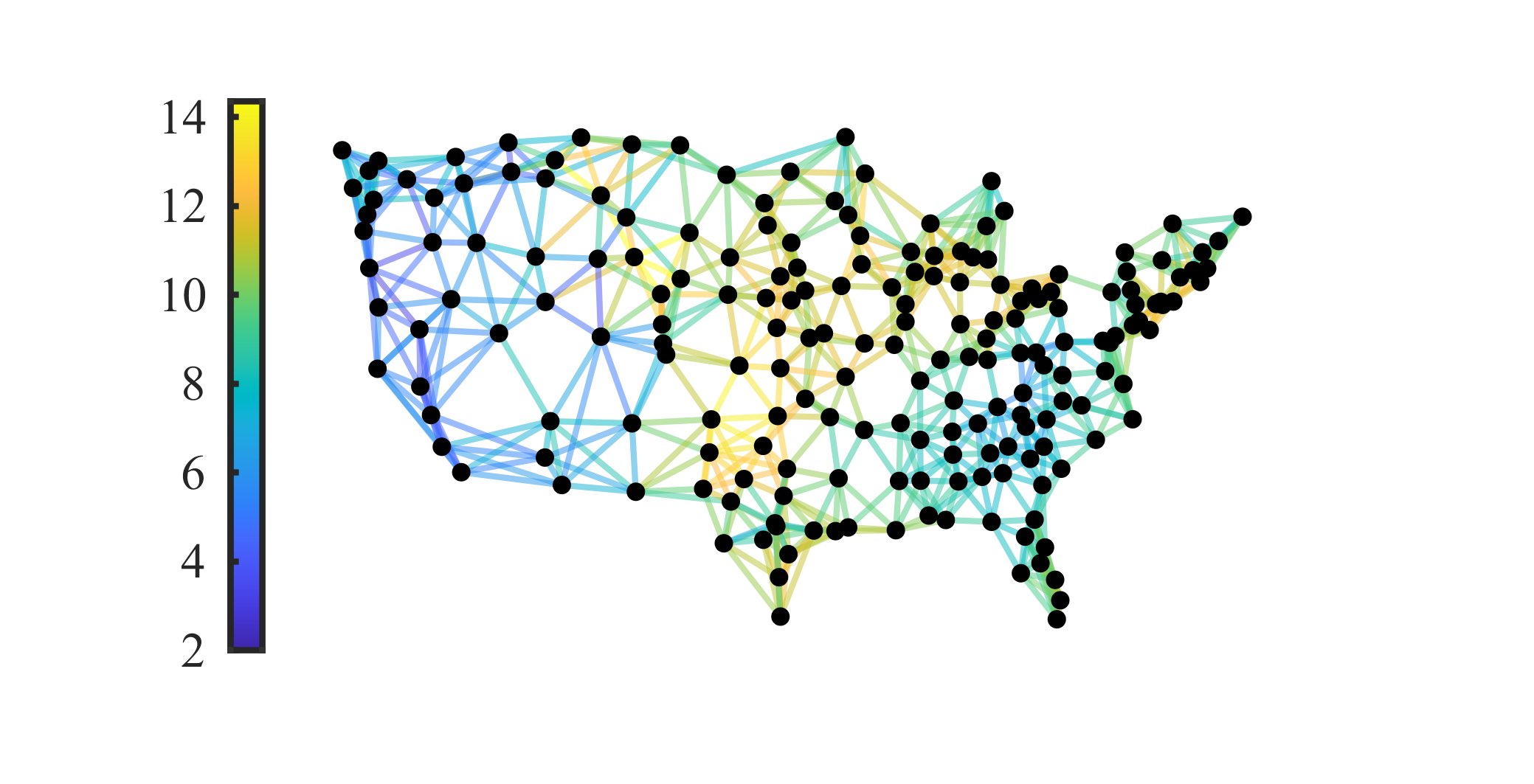}
                   \vspace{-30pt}
         \caption*{$t=12$}
     \end{subfigure}
        \caption{A graph with time-varying wind speed on the edges.}
        \label{fig_wind}
\end{figure}

In Graph Signal Processing (GSP), the Graph Least Mean Squares (GLMS) algorithm is a well-known adaptive method for online estimation of time-varying node signals under Gaussian noise \cite{bib_LMS}.
A time-varying node signal, denoted as $\boldsymbol{x}_n[t]$, is the function values defined on the graph nodes. 
Several adaptive GSP techniques have been introduced to enhance aspects such as convergence speed and robustness to impulsive non-Gaussian noise \cite{bib_NLMS, yan_2022_sign, bib_LMP, yan_2022_NLMP, li2023robust}. 
To implement the GFT, we need to conduct the eigen decomposition on $\mathbf{L}$: 
\begin{equation}
    \mathbf{L=U\Lambda U}^{T}, 
\end{equation}
where $\mathbf{U}$ is the eigenvector matrix and $\mathbf{\Lambda} = $ diag $(\lambda_1 ... \lambda_N)$ is the eigenvalue matrix.
To give a sense of low and high frequencies, the eigenvalue-eigenvector pairs are sorted in increasing order \cite{Ortega_2018}. 
The node signal can be transformed to the spectral domain by the forward GFT $\boldsymbol{s}_n[t] = \mathbf{U}^{T}\boldsymbol{x}_n[t]$. 
The inverse transform is $\boldsymbol{x}_n[t] = \mathbf{U}\boldsymbol{s}_n[t]$. 
Given a graph filter $\mathbf{\Sigma}$, the most basic GSP spectral filtering operation is \begin{equation}
    \boldsymbol{x}_n'[t] = \mathbf{U\Sigma U}^{T}\boldsymbol{x}_n[t].
\end{equation}

\section{Methodology}
\label{sec_method}
\begin{figure}[h]
     \centering
     \begin{subfigure}{0.45\textwidth}
         \centering
         \includegraphics[width=0.9\textwidth]{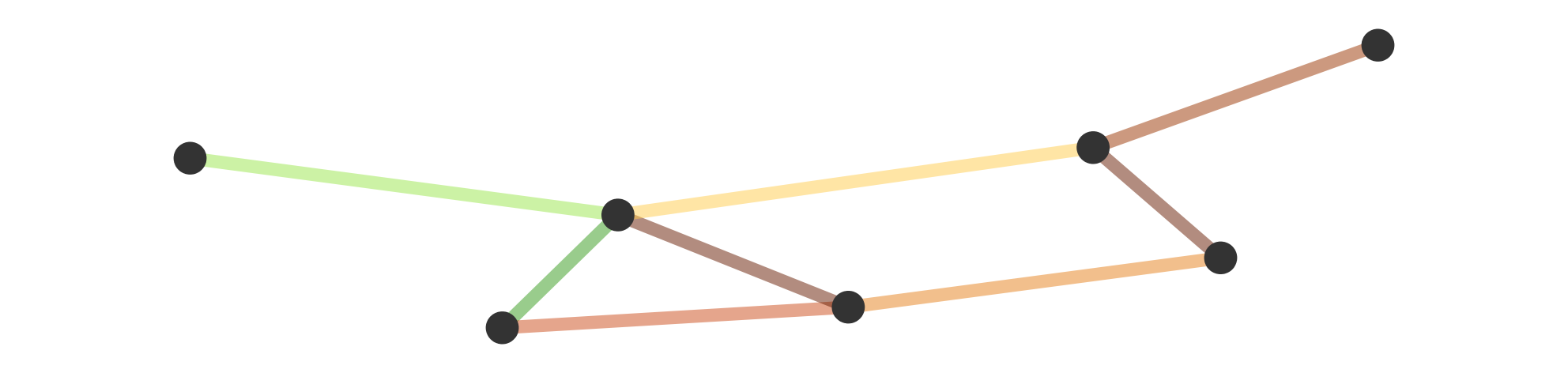}
         \caption*{$\mathcal{G}$}    
     \end{subfigure}      
     \begin{subfigure}{0.45\textwidth}
         \centering
         \includegraphics[width=0.9\textwidth]{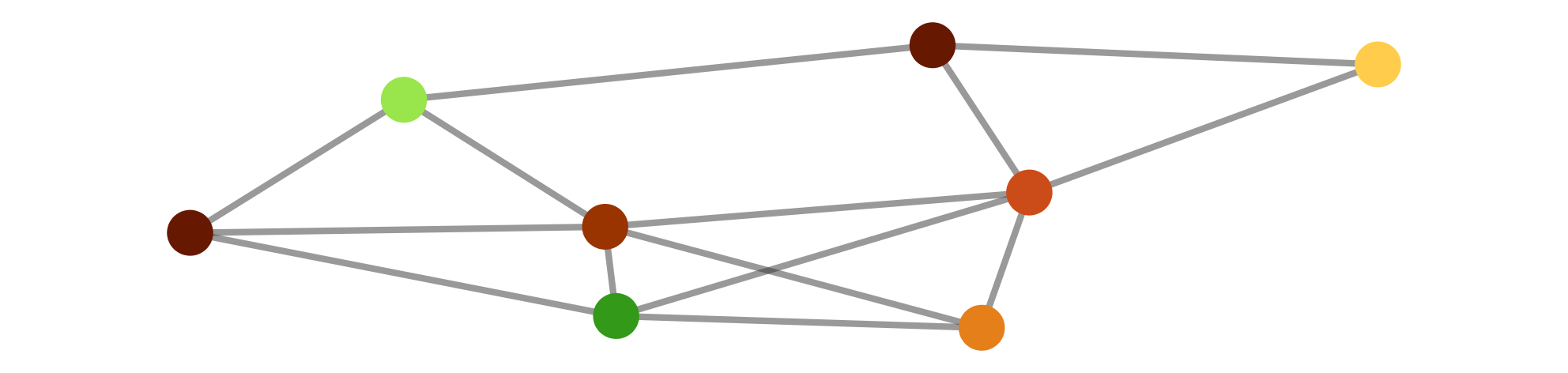}
         \caption*{$\mathcal{G}_{LG}$}
     \end{subfigure}
        \caption{A graph $\mathcal{G}$ and its line graph $\mathcal{G}_{LG}$.}
        \label{fig_line}
\end{figure}

Given a graph $\mathcal{G}$, its edge-to-vertex dual is known as the Line Graph of $\mathcal{G}$, denoted as $\mathcal{G}_{LG}$. 
In this paper, we assume the topology of $\mathcal{G}$ is static, but the signals on the edges are time-varying.
The adjacency matrix of $\mathcal{G}_{LG}$ can be constructed using the (oriented) node-to-edge incidence matrix $\mathbf{B}$:
\begin{equation}
    \mathbf{A}_{LG} = \text{abs}(\mathbf{B}^T\mathbf{B})-2\mathbf{I},
    \label{line_graph_adjacency}
\end{equation}
where $\mathbf{I}$ is the identity matrix. 
In other words, to connect the nodes of $\mathcal{G}_{LG}$, an edge is placed between two nodes of $\mathcal{G}_{LG}$ if their corresponding edges in $\mathcal{E}$ are connected to the same node in $\mathcal{G}$. 
We can follow the definitions above to form the Laplacian matrix $\mathbf{L}_{LG}$ for the Line Graph. 
The edge signals $\boldsymbol{x}_e[t]$ of $\mathcal{G}$ can be treated as the node signals  $\boldsymbol{x}_n[t]$ of $\mathcal{G}_{LG}$ if we transform $\mathcal{G}$ to $\mathcal{G}_{LG}$; all the GSP techniques such as GFT, convolution, filtering, and sampling remains unchanged without the need of defining edge-specific techniques. 
A graph with edges signals and its line graph is demonstrated in Fig.~\ref{fig_line}.

Assuming that the edge signals are noisy, we can model the noise on the nodes $\boldsymbol{\omega}_n[t]$ by Gaussian distributions with zero mean. 
Missing node observations can be modeled using a masking matrix $\mathbf{M}$, where the $i^{th}$ diagonal is an indicator of whether the $i^{th}$ node is missing or not. 
The signal on the nodes with missing value and noise at time $t$ is then 
\begin{equation}
    \boldsymbol{y}_n[t] = \mathbf{M}(\boldsymbol{x}_n[t]+\boldsymbol{\omega}_n[t]).
\end{equation}
Notice that the missing nodes can also follow a predefined sampling strategy in order to maximize the desired properties of graphs or to enforce spatial domain sparsity \cite{di2018adaptive_sampling}.
In LGLMS, we will adopt a similar data model for the edge signals:
\begin{equation}
    \boldsymbol{y}_e[t] = \mathbf{M}(\boldsymbol{x}_e[t]+\boldsymbol{\omega}_e[t]).
\end{equation}
Assuming that the signal of interest is a time-varying edge signal $\boldsymbol{x}_e[t]$, if the edges of $\mathcal{G}$ are mapped to the nodes of $\mathcal{G}_{LG}$, we can process the edge signal as node signals. 
In most cases, a graph shift operator is required in the GSP algorithm, which can be either the adjacency matrix or the graph Laplacian matrix \cite{Ortega_2018}. 
Choosing $\mathbf{L}_{LG}$ as the graph shift, spectral domain operations on the edges can be defined again using the GFT on $\mathcal{G}_{LG}$:
\begin{equation}
    \mathbf{L}_{LG} = \mathbf{U}_{LG}\mathbf{\Lambda}_{LG}\mathbf{U}^{T}_{LG}.
    \label{LG_GFT}
\end{equation}
Afterward, we can process the time-varying signals using the following model:
\begin{equation}
    \boldsymbol{x}_e[t+1] = \boldsymbol{x}_e[t]+\Delta_e[t],
    \label{model}
\end{equation}
where $\Delta_e[t]$ is the change in the edge signal that leads $\boldsymbol{x}_e[t]$ to $\boldsymbol{x}_e[t+1]$.
Then, we define a spectral domain filter $\mathbf{\Sigma}_{LG}$ based on assumptions of the edge signals, such as smoothness or bandlimitedness.
To obtain $\Delta_e[t]$,  we can rely on minimizing the following $l_2$-norm optimization problem similar to what is seen in the GLMS \cite{bib_LMS}:
\begin{equation}
    J(\hat{\boldsymbol{x}}_e[t])=\mathbb{E}\left\|\boldsymbol{y}_e[t]-\mathbf{M}\mathbf{U}_{LG}\mathbf{\Sigma}_{LG}\mathbf{U}^{T}_{LG}\hat{\boldsymbol{x}}_e[t]\right\|_2^2.
    \label{cost}
\end{equation}
\begin{figure}
    \centering
    \includegraphics[width=\linewidth]{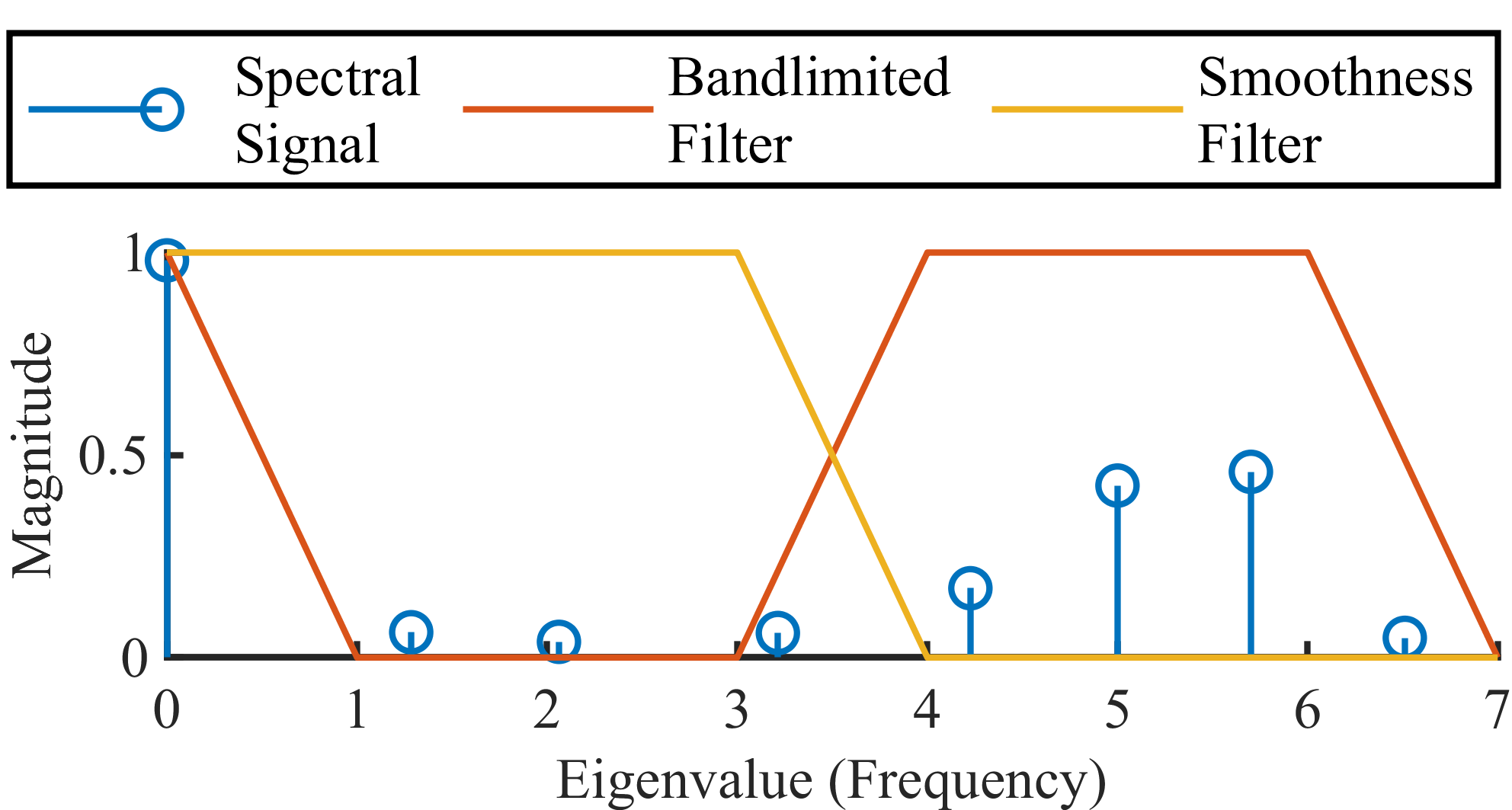}
    \caption{Comparing bandlimited and smoothness-based filter.}
    \label{fig_filter}
\end{figure}
In LGLMS, we will adopt a bandlimited design for $\mathbf{\Sigma}_{LG}$, which is based on a spectrum similar to the ground truth data. 
Unlike conventional smoothness-based low-pass filters, which impose restrictive assumptions on signal variation, the bandlimited approach enables selective frequency preservation, allowing for more representational power over the key spectral content of edge signals.
A comparison of a bandlimited filter with a smoothness-based filter is in Fig.~\ref{fig_filter}.
In practice, the required spectral information can be derived from historical observations or approximated from noisy measurements, ensuring adaptability in real-world applications.
The bandlimitedness fundamentally reaches beyond smoothness, which emerges as a special case within the broader bandlimited paradigm. 
By breaking the constraints of traditional smoothness priors, LGLMS overcomes the inherent limitations—particularly its inability to retain critical high-frequency components essential for capturing dynamic edge signal variations.
Based on bandlimitedness  of the edge signal, we can obtain 
\begin{equation}
        \Delta_e[t] = \frac{\partial f(\hat{\boldsymbol{x}}_e[t])}{\partial\hat{\boldsymbol{x}}_e[t]}        =-2\mathbf{U}_{LG}\mathbf{\Sigma}_{LG}\mathbf{U}^{T}_{LG}\mathbf{M}(\boldsymbol{y}_e\left[t\right]-\hat{\boldsymbol{x}}_e\left[t\right]).
        \label{delta}
\end{equation}

Plugging \eqref{delta} into \eqref{model}, the update function of the LGLMS algorithm gives the next step edge signal prediction $\hat{\boldsymbol{x}}_e[t+1]$:
\begin{equation}
    \hat{\boldsymbol{x}}_e\left[t+1\right]=\hat{\boldsymbol{x}}_e\left[t\right]+\alpha \mathbf{U}_{LG}\mathbf{\Sigma}_{LG}\mathbf{U}^{T}_{LG}\mathbf{M}(\boldsymbol{y}_e\left[t\right]-\hat{\boldsymbol{x}}_e\left[t\right]),
    \label{LGLMS_update}
\end{equation}
where $\alpha$ is the step size. 
For simplicity, we assume that the graph  $\mathcal{G}$ is fixed, but we should point out that the change in the graph structure does not change the definitions of any GSP techniques, and the Line Graph transformation remains valid.

Here is an intuitive explanation of the complete procedure of using the LGLMS to process edge signals. 
Given a graph $\mathcal{G}$, begin the LGLMS by constructing the Line Graph $\mathcal{G}_{LG}$ as seen in \eqref{line_graph_adjacency}. 
To conduct spectral domain operations, we define the necessary GFT components not using the original graph $\mathcal{G}$ but using the Line Graph $\mathcal{G}_{LG}$ using \eqref{LG_GFT} and define a spectral domain filter $\mathbf{\Sigma}_{LG}$ (preferably a bandlimited close to the ground truth). 
While there are noisy and missing edge signal observations $\boldsymbol{y}_e[t]$, iteratively execute the LGLMS algorithm shown in \eqref{LGLMS_update}. 
By embedding edge signals into the node space of the Line Graph, LGLMS enables an adaptive learning process that inherently captures the key spectral characteristics of the edge signals.
The pseudocode for implementing LGLMS online spatiotemporal estimation of edge signals is illustrated in Algorithm~\ref{algorithm_line}.

\begin{algorithm}
\caption{LGLMS: online estimation of edge signals}
\begin{algorithmic}[1]
    \State Generate the Line Graph $\mathcal{G}_{LG}$ from $\mathcal{G}$ using \eqref{line_graph_adjacency}
    \State Define the GFT of $\mathcal{G}_{LG}$ 
    \State Define bandlimited filter $\mathbf{\Sigma}_{LG}$ using \eqref{LG_GFT}
    \While {There is new edge signal observation $\boldsymbol{y}_e[t]$}
        \State {\bf{Input:}} the current edge signal observation $\boldsymbol{y}_e[t]$
        \State Execute the signal prediction strategy in \eqref{LGLMS_update}
        \State {\textbf{Output:}} Estimated edge signals $\hat{\boldsymbol{x}}_e[t]$
    \EndWhile
\end{algorithmic}
\label{algorithm_line}
\end{algorithm}

The choice of $l_2$-norm optimization not only ensures a mathematically tractable solution but also facilitates an efficient and scalable implementation, making it a foundational choice for designing adaptive algorithms in the online estimation of time-varying signals. 
The LGLMS is the parallel of the classical adaptive LMS algorithm on the graph edges.
What LGLMS differs from classical LMS is that LGLMS relies on the bandlimitedness of the edge signals, reducing the update complexity of dynamically changing the filter weights. 
LGLMS can interpolate missing edge observations due to the operation performed by $\mathbf{U}_{LG}\mathbf{\Sigma}_{LG}\mathbf{U}^{T}_{LG}\mathbf{M}(\boldsymbol{y}_e[t] - \hat{\boldsymbol{x}}_e[t])$ is an edge diffusion.
The adaptive GLMS algorithm defined on the graph nodes is stable and converges in mean mean-squared sense under steady state estimation for the selection of $\Sigma_{LG}$ and $\alpha$ satisfying the following condition:
\begin{equation}
    \|\alpha\mathbf{M}\mathbf{U}_{LG}\mathbf{\Sigma}_{LG}\mathbf{U}^{T}_{LG}\|^2_2  \leq 1.
\end{equation}
This convergence behavior of LGLMS is similar to what can be seen in other adaptive GSP methods \cite{bib_LMS, bib_LMP, yan_2022_sign}.

LGLMS operates on the edges of graph $\mathcal{G}$ that are embedded into the node space of the Line Graph $\mathcal{G}_{LG}$, and all operations are defined using the node space of $\mathcal{G}_{LG}$.
Compared with conventional GSP methods, under fixed graph topology, the only additional calculation is one matrix multiplication and one matrix addition in \eqref{line_graph_adjacency}, and the majority of the computational complexity will be dominated by the algorithm update \eqref{LGLMS_update}. 
When the graph is dynamic, the eigendecomposition in the GFT will dominate the GSP. 
In other words, the Line Graph transform does not introduce additional computational complexity to LGLMS when compared against GSP algorithms that are deployed on the graph nodes.
The Line Graph transformation converted the edge signals from edge space to node space, letting LGLMS unify well-developed GSP techniques to process signals with classical adaptive filters on the graph edges, fundamentally distinguishing itself from conventional GSP methods, which are constrained to node-signal processing. 

\section{Experiment and Discussion}
LGLMS is particularly well-suited for applications involving spatial-temporal edge dynamics, such as real-time traffic flow prediction and dynamic environmental monitoring.
The LGLMS algorithm is tested on two different sets of data. 
The first data is the Sioux Falls network with $N_n =  24$ nodes and $N_e = 38$ edges. 
The Sioux Falls network is a traffic network based on real-world road maps, where the edges are the roads and the edge signals are time-invariant traffic flows on roads \cite{transportation}. 
We simulated time-varying behavior from the given time-invariant edge signal by multiplying it with a summation of different multivariate sinusoidal signals. 
The second data, the U.S. meteorological data, is a graph with $N_n = 197$ nodes and $N_e = 818$ edges \cite{bib_weather_dataset}. 
Each weather station is a node on the graph where the stations are connected to their near geographical neighbors using a distance-based metric, as seen in \cite{bib_NLMS}. 
In our experiment, we built the model using the dataset as follows: for each edge, the signal on it is formed by taking the average of the real meteorological recordings on the two weather stations on that edge. 
Hourly temperature and hourly wind speed are selected as the two target features because they align with the smoothness assumption: neighboring nodes have similar values, and the readings are correlated across adjacent nodes.
A visualization of the meteorological graph with the time-varying wind speed is shown in Fig.~\ref{fig_wind}.  
For both datasets, we will be adding Gaussian noise and setting only 2/3 of the edges as observed edge signals using two types of observation masks. 
The first observation mask is created for each experiment run, in which the random missing observation is aimed at mimicking the missing data measurements in the real world.
The second observation mask is to create a greedy smoothness-based sampling set using the sampling approach seen in \cite{yanagiya2022edge_sampling}: create a subset of edges that maximizes the low-frequency content (edge signal smoothness assumption) to be the observation mask. 
We run all algorithms for the Sioux Falls network based on two different missing edge signal scenarios.
For the meteorological dataset, we only use the random sampling strategy for both features since the high number of edges makes it difficult to realize the greedy sampling approach seen in \cite{yanagiya2022edge_sampling}.

The LGLMS is compared against two baselines. 
The first baseline (denoted as Spectral) is a non-adaptive filter that also projects edge signals onto the Line graph, similar to the LGLMS: $\hat{\boldsymbol{x}}_e[t+1] = \mathbf{U}_{LG}\mathbf{\Lambda}_{LG}\mathbf{U}^{T}_{LG}{\boldsymbol{y}}_e[t]$. 
The second baseline is the Simplicial Convolution (SC): $\hat{\boldsymbol{x}}_e[t+1] = \theta \mathbf{L}_{l}\hat{\boldsymbol{x}}_e[t]+\gamma\mathbf{L}_{u}\hat{\boldsymbol{x}}_e[t]+\xi\hat{\boldsymbol{x}}_e[t]$, where $\mathbf{L}_{l}$ is the lower Hodge Laplacian, $\mathbf{L}_{u}$ is the upper Hodge Laplacian, $\theta, \gamma$, and $\xi$ are the parameters as defined in \cite{Yang_2022_Simplicial}. 
We implement each of the 3 tested methods with two types of filters, giving us 6 different algorithms in total.  
The first filter, denoted with subscript LP, is a low-pass filter based on the smoothness assumption of the edge signals  \cite{yanagiya2022edge_sampling}. 
The second filter, denoted with subscript BL, is a bandlimited filter based on the bandlimited assumption of the edge signals. 
To measure the prediction accuracy, we calculate the normalized mean square error between the predicted value $\hat{\boldsymbol{x}}_e[t]$ and the ground truth $\boldsymbol{x}_e[t]$ at each time instance of the online prediction:
$\text{NMSE}[t] = \frac{1}{N_r}\sum_{i=1}^{N_e}{\frac{(x_i[t]-\hat{x}_{i}[t])^2}{(x_i[t])^2}}$, 
where $N_r$ is the number of experiment runs, $\hat{x}_{i}[t]$ is the predicted signal on the $i^{th}$ edge at time $t$, and $x_i[t]$ is the ground truth signal on the $i^{th}$ edge at time $t$. 

\begin{figure}
    \centering
    \includegraphics{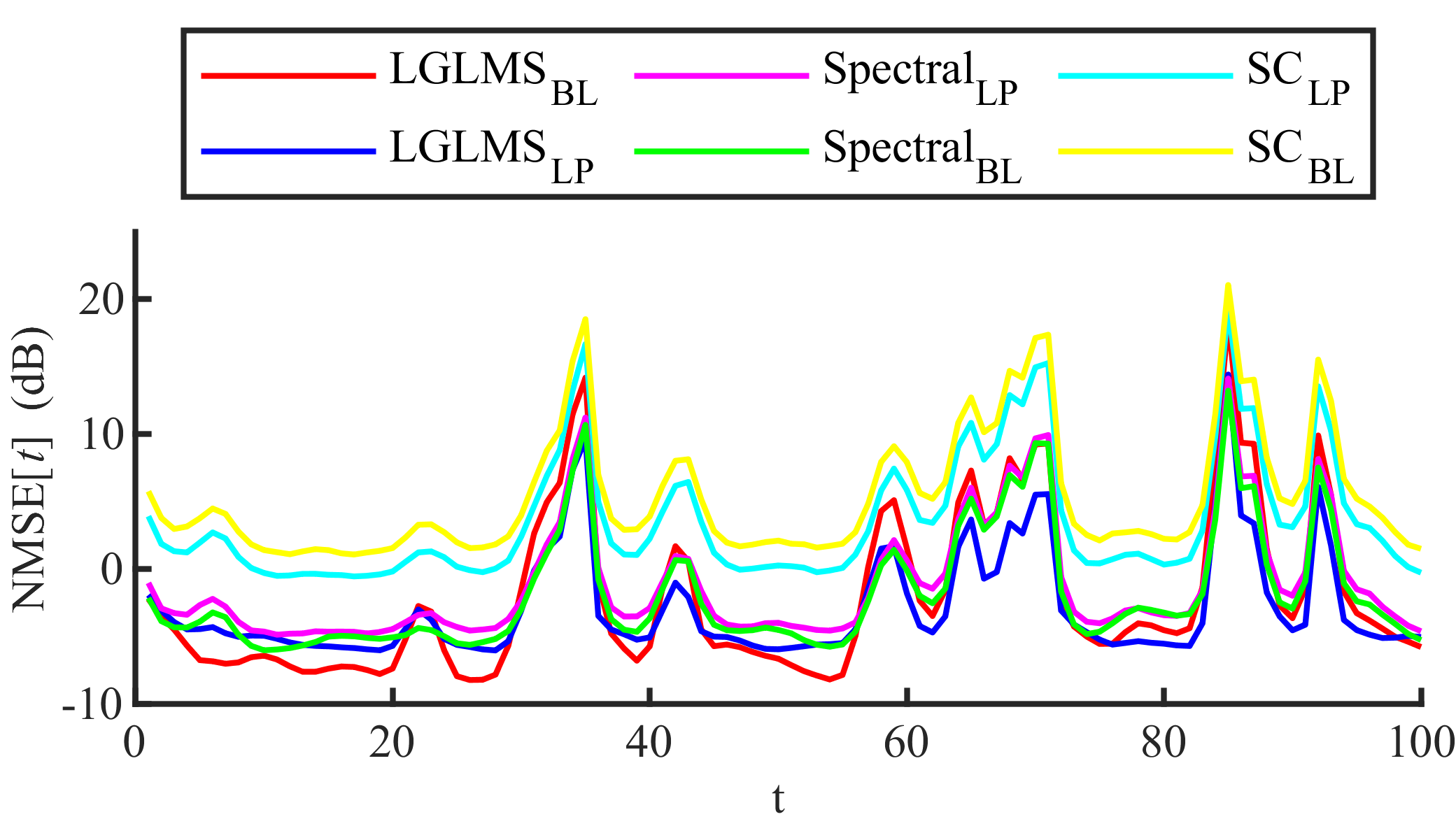}
    \caption{The NMSE on the Sioux Falls Network using a low-pass sampling strategy.}
    \label{fig_MSE_low_pass}
\end{figure}

\begin{figure}
    \centering
    \includegraphics{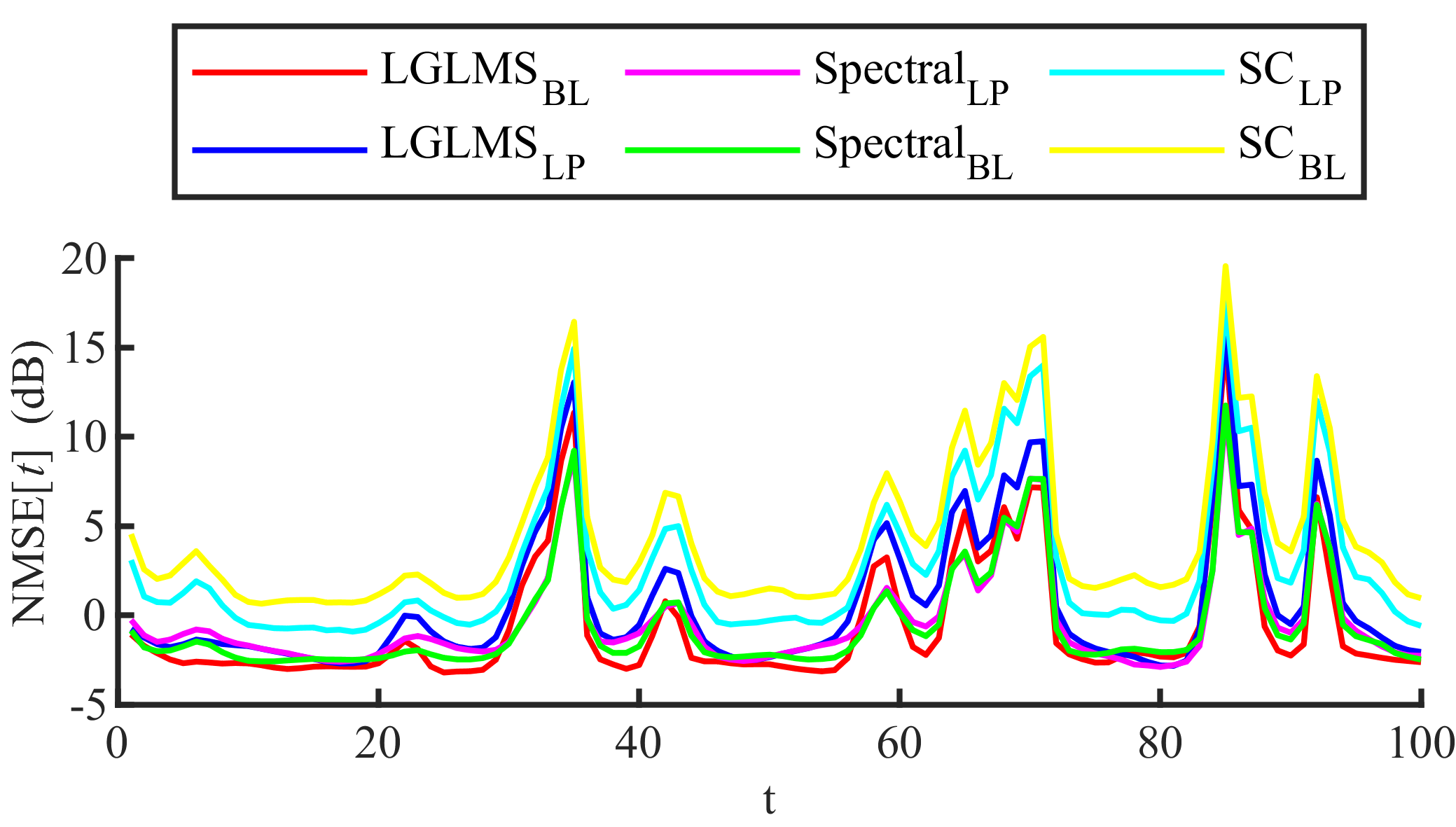}
    \caption{The NMSE on the Sioux Falls Network using a random sampling strategy.}
    \label{fig_MSE_rand}
\end{figure}

On the Sioux Falls Network, the resulting NMSE$[t]$ of using a smoothness-based sampling strategy is shown in Fig.~\ref{fig_MSE_low_pass}, and the resulting NMSE$[t]$ of using a random sampling strategy is shown in Fig.~\ref{fig_MSE_rand}.
Analyzing Fig.~\ref{fig_MSE_low_pass} for the smoothness-based sampling case, we see that for both the LGLMS$_\text{BL}$ and the LGLMS$_\text{LP}$, LGLMS have relatively lower NMSE$[t]$ compared with other baselines.
The performance of LGLMS$_\text{BL}$ and LGLMS$_\text{LP}$ is similar because even though the smoothness-based sampling is in favor of the low-pass filter, we can still achieve a low-pass effect using the bandlimited filter.
Looking at the random missing case in Fig.~\ref{fig_MSE_rand}, we can see that the LGLMS$_\text{BL}$ makes predictions that result in lower NMSE$[t]$ for most of the time instances compared to the other baseline methods. 
Analyzing both Fig.~\ref{fig_MSE_low_pass} and Fig.~\ref{fig_MSE_rand}, the edge signals onto the nodes of the Line Graph can indeed give GSP algorithms the ability to process signals on the edges. 
We notice that in Fig.~\ref{fig_MSE_rand}, LGLMS performs worse when using a low-pass filter than when using a bandlimited filter. 
The reason behind the better performance of a bandlimited filter over a low-pass filter is that, under random missing of the edge signals with Gaussian noise, the observed signals are no longer guaranteed to be smooth.
However, the low-pass filter has the underlying assumption of the smoothness of the signal. 
This makes a properly defined bandlimited filter that is closer to the ground truth spectrum of the signal perform better. 
Another factor that contributes to the good performance of the LGLMS is its adaptive update strategy, allowing it to capture the time-varying dynamics of the time-varying edge signal. 
This simple yet effective adaptive update scheme is unique to adaptive filters but is lacking in the other baselines.

\begin{figure}
    \centering
    \includegraphics{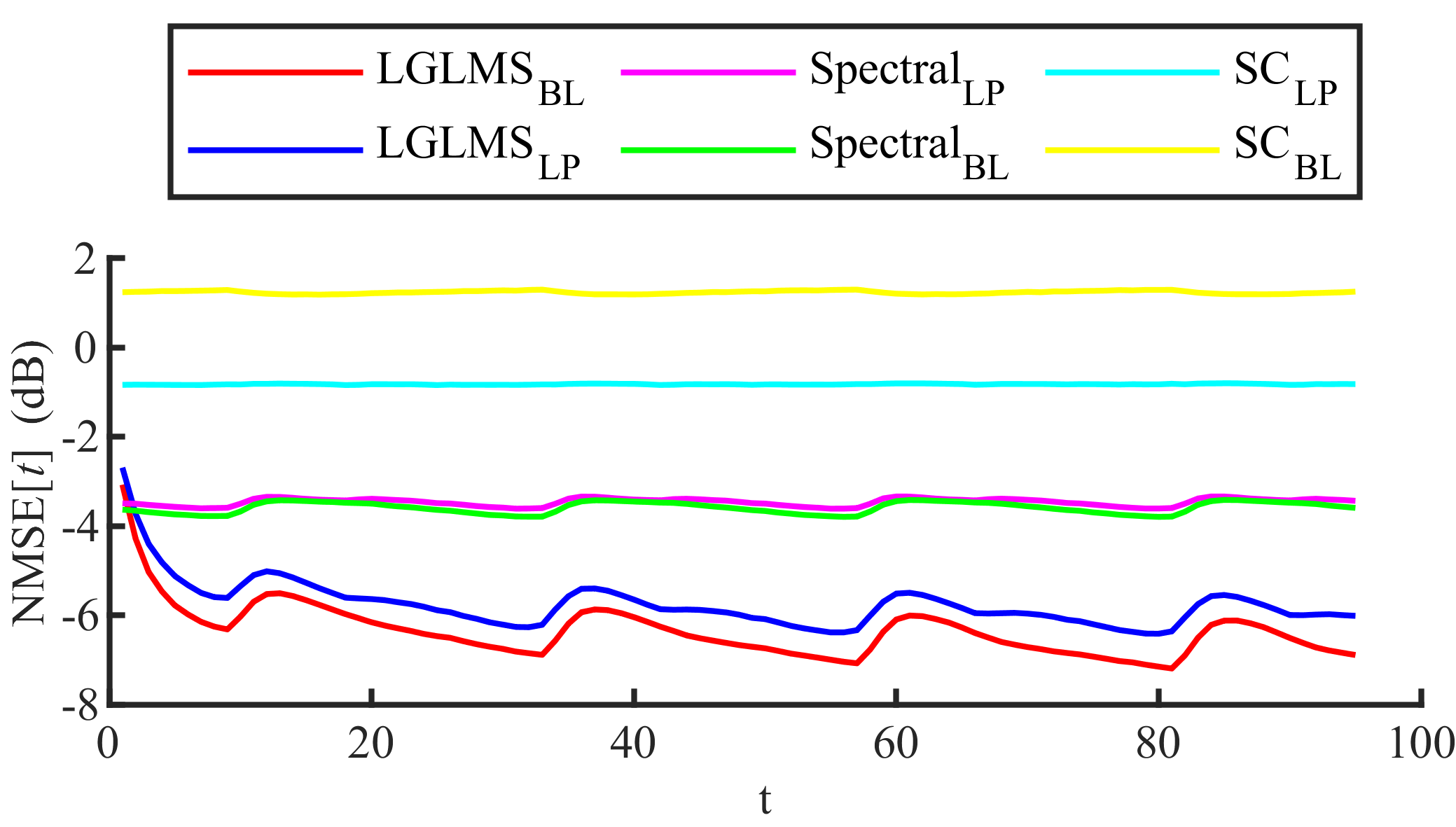}
    \caption{The NMSE on the temperature prediction using a random sampling strategy.}
    \label{fig_MSE_temp}
\end{figure}

\begin{figure}
    \centering
    \includegraphics{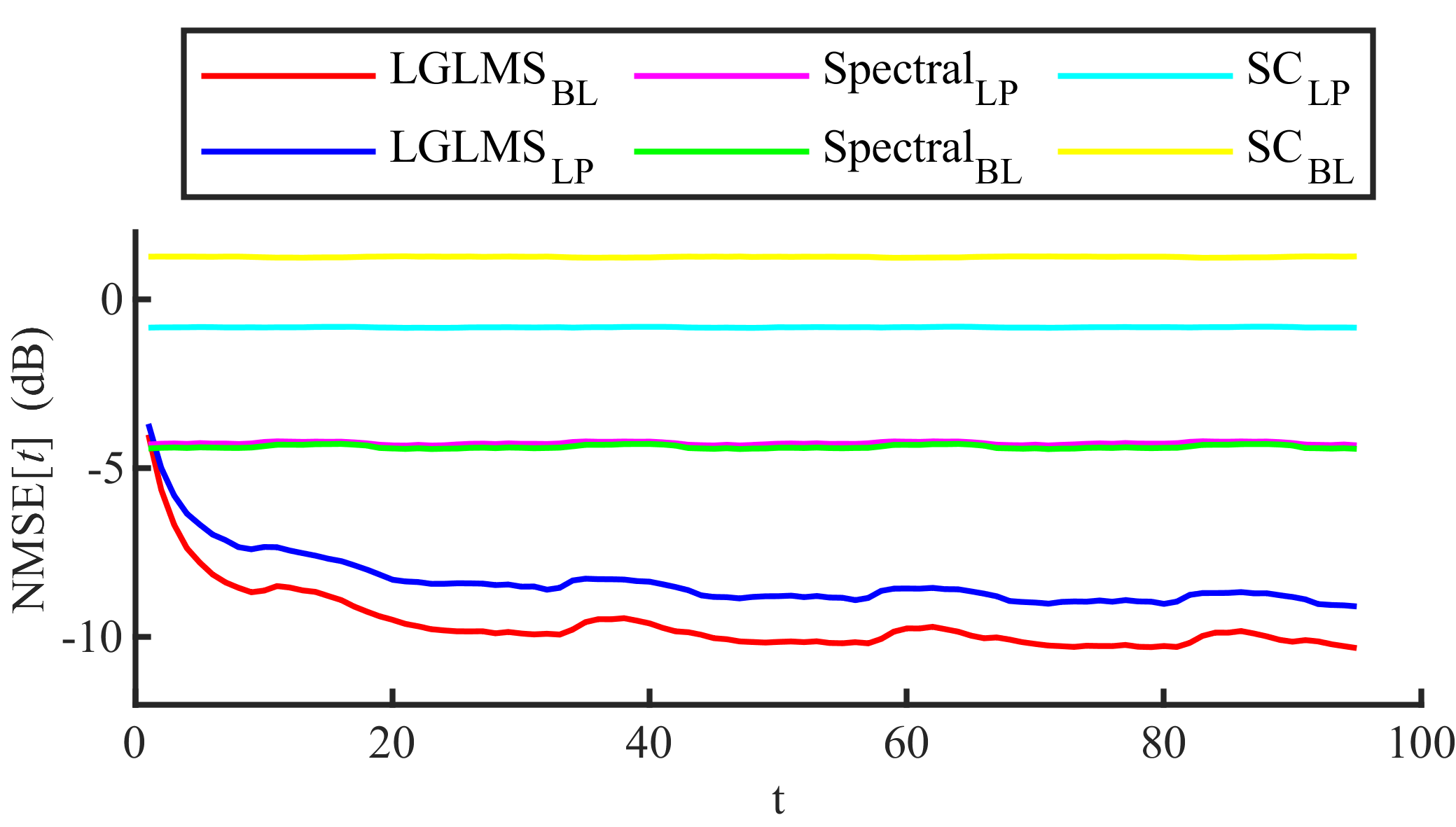}
    \caption{The NMSE on the wind speed prediction using a random sampling strategy.}
    \label{fig_MSE_wind}
\end{figure}

For the meteorological network, the NMSE$[t]$ of the temperature predictions is shown in Fig.~\ref{fig_MSE_temp}, and the NMSE$[t]$ of the wind speed predictions is shown in Fig.~\ref{fig_MSE_wind}. 
Inspecting Fig.~\ref{fig_MSE_temp} and  Fig.~\ref{fig_MSE_wind}, both LGLMS$_\text{BL}$ and LGLMS$_\text{LP}$ have lower NMSE$[t]$ than the other baselines. 
The low NMSE$[t]$ on the meteorological network again indicates the effectiveness of the LGLMS at the online prediction of time-varying edge signals. 
The consistently lower NMSE$[t]$ of LGLMS$_\text{BL}$ compared with LGLMS$_\text{LP}$ indicates that the bandlimited filter is a more suitable filter choice.
Additionally, the mean run time of each prediction step is recorded at 0.1732 seconds for experiments on the meteorological network for predicting hourly weather changes, indicating that LGLMS operates within realistic deployment time constraints for real-time applications.
It should be noticed that the meteorological network ($N_e = 818$ edges) has a larger scale than the Sioux Falls network ($N_e = 38$ edges), and LGLMS performs well on both datasets, indicating that LGLMS has scalability potential for larger datasets.

\section{Conclusion and Future Works}
LGLMS offers a practical and efficient solution for learning time-varying edge signals under noise and missing observations, seamlessly integrating classical adaptive filters with GSP on graph edges through edge signal embedding. To evaluate its effectiveness, we conduct experiments on graphs of varying sizes and applications; results confirm that LGLMS delivers accurate and robust online prediction of time-varying edge signals, demonstrating its reliability across diverse real-world conditions.
This novel formulation preserves the structural relationships of edge signals while unlocking the full potential of adaptive GSP on graph edges, enabling a cohesive and computationally efficient approach to edge-based signal processing and machine learning on graphs.

Our future work will explore the incorporation of message passing techniques in the proposed LGLMS \cite{yan2025graph, yan2024llm, qin2025llm}. 
Additionally, we aim to investigate the integration of GSP principles with Large Language Models (LLMs) to harness the reasoning capabilities of powerful LLMs \cite{yan2024llm, qin2025llm}. 
We will also address edge signals under impulsive noise, which is crucial for improving robustness and accuracy in real-world applications \cite{kuruoglu1997new}.

\section{Acknowledgments}
This work is supported by Shenzhen Science and Technology Innovation Commission under Grant JCYJ20220530143002005, Shenzhen Ubiquitous Data Enabling Key Lab under Grant ZDSYS20220527171406015, and Tsinghua Shenzhen International Graduate School Start-up fund under Grant QD2022024C.
\section{References}
\vspace{-20pt}
\renewcommand{\refname}{}
\bibliographystyle{IEEEbib}
\bibliography{refs.bib}

@article{yan2024signal,
  title={Signal processing over time-varying graphs: A systematic review},
  author={Yan, Yi and Hou, Jiacheng and Song, Zhenjie and Kuruoglu, Ercan Engin},
  journal={arXiv},
  year={2024}
}

@article{li2023robust,
  title={Robust Recovery for Graph Signal via l0-Norm Regularization},
  author={Li, Xiao Peng and Yan, Yi and Kuruoglu, Ercan Engin and So, Hing Cheung and Chen, Yuan},
  journal={IEEE Signal Processing Letters},
  volume={30},
  pages={1322--1326},
  year={2023},
  publisher={IEEE}
}

@article{yan2025binarized,
  title={Binarized simplicial convolutional neural networks},
  author={Yan, Yi and Kuruoglu, Ercan Engin},
  journal={Neural Networks},
  volume={183},
  pages={106928},
  year={2025},
  publisher={Elsevier}
}

@article{yan2025graph,
  title={Graph Signal Adaptive Message Passing},
  author={Yan, Yi and Peng, Changran and Kuruoglu, Ercan E},
  journal={IEEE Signal Processing Letters},
  year={2025},
  publisher={IEEE}
}

@article{yan2024llm,
  title={{LLM} online spatial-temporal signal reconstruction under noise},
  author={Yan, Yi and Qin, Dayu and Kuruoglu, Ercan Engin},
  journal={arXiv preprint arXiv:2411.15764},
  year={2024}
}

@inproceedings{qin2025llm,
  title={{LLM}-based Online Prediction of Time-varying Graph Signals (Student Abstract)},
  author={Qin, Dayu and Yan, Yi and Kuruoglu, Ercan Engin},
  booktitle={Proceedings of the AAAI Conference on Artificial Intelligence},
  volume={39},
  number={28},
  pages={29472--29474},
  year={2025}
}

@InProceedings{wang_2024_understanding_heterophily,
  title = 	 {Understanding Heterophily for Graph Neural Networks},
  author =       {Wang, Junfu and Guo, Yuanfang and Yang, Liang and Wang, Yunhong},
  booktitle = 	 {ICML},
  pages = 	 {50489--50529},
  year = 	 {2024},
  editor = 	 {Salakhutdinov, Ruslan and Kolter, Zico and Heller, Katherine and Weller, Adrian and Oliver, Nuria and Scarlett, Jonathan and Berkenkamp, Felix},
  volume = 	 {235},
  series = 	 {Proceedings of Machine Learning Research},
  publisher =    {PMLR}
}

@INPROCEEDINGS{Marinucci_2024_cell_complex,
  author={Marinucci, Lorenzo and Battiloro, Claudio and Lorenzo, Paolo Di},
  booktitle={2024 32nd European Signal Processing Conference (EUSIPCO)}, 
  title={Topological Adaptive Learning over Cell Complexes}, 
  year={2024},
  volume={},
  number={},
  pages={832-836},
  doi={10.23919/EUSIPCO63174.2024.10714988}}

@article{luan2022revisiting_heterophily,
  title={Revisiting heterophily for graph neural networks},
  author={Luan, Sitao and Hua, Chenqing and Lu, Qincheng and Zhu, Jiaqi and Zhao, Mingde and Zhang, Shuyuan and Chang, Xiao-Wen and Precup, Doina},
  journal={NeruIPS},
  volume={35},
  pages={1362--1375},
  year={2022}
}

@article{Schaub_2021_higher_order_networks,
title = {Signal processing on higher-order networks: Livin’ on the edge... and beyond},
journal = {Signal Processing},
volume = {187},
pages = {108149},
year = {2021},
author = {Michael T. Schaub and Yu Zhu and Jean-Baptiste Seby and T. Mitchell Roddenberry and Santiago Segarra}
}

@ARTICLE{Leus_2023_GSP,
  author={Leus, Geert and Marques, Antonio G. and Moura, José M.F. and Ortega, Antonio and Shuman, David I},
  journal={IEEE Signal Processing Magazine}, 
  title={Graph Signal Processing: History, development, impact, and outlook}, 
  year={2023},
  volume={40},
  number={4},
  pages={49-60},
  doi={10.1109/MSP.2023.3262906}}

@inproceedings{Fan_2023_social,
author = {Fan, Wenqi and Ma, Yao and Li, Qing and He, Yuan and Zhao, Eric and Tang, Jiliang and Yin, Dawei},
title = {Graph Neural Networks for Social Recommendation},
year = {2019},
booktitle = {WWW},
series = {WWW}
}

@ARTICLE{Stefania_2024_cell_complex,
  author={Sardellitti, Stefania and Barbarossa, Sergio},
  journal={IEEE Trans. Signal Process.}, 
  title={Topological Signal Processing Over Generalized Cell Complexes}, 
  year={2024},
  volume={72},
  number={},
  pages={687-700}}

@INPROCEEDINGS{Roddenberry_cell_complex_2022,
  author={Roddenberry, T. Mitchell and Schaub, Michael T. and Hajij, Mustafa},
  booktitle={ICASSP}, 
  title={Signal Processing On Cell Complexes}, 
  year={2022},
  volume={},
  number={},
  pages={8852-8856}}

@article{dong2020graph,
  title={Graph signal processing for machine learning: A review and new perspectives},
  author={Dong, X. and Thanou, D. and Toni, L. and Bronstein, M. and Frossard, P.},
  journal={IEEE Signal Process. Mag.},
  volume={37},
  number={6},
  pages={117--127},
  year={2020},
  publisher={IEEE}
}

@ARTICLE{bib_weather_dataset,
author   = {Palecki, M. and Durre, I. and Applequist, S. and Arguez, A and Lawrimore, J. },
journal = {NOAA National Centers for Environmental Information},
title     = {{U.S.} Climate Normals 2020: {U.S.} Hourly Climate Normals (1991-2020)},
year = {2020}}

@inproceedings{yanagiya2022edge_sampling,
  title={Edge sampling of graphs based on edge smoothness},
  author={Yanagiya, K. and Yamada, K. and Katsuhara, Y. and Takatani, T. and Tanaka, Y.},
  booktitle={ICASSP},
  pages={5932--5936},
  year={2022},
  organization={IEEE}
}

@article{di2018adaptive_sampling,
  title={Adaptive graph signal processing: Algorithms and optimal sampling strategies},
  author={D. Lorenzo, P. and Banelli, P. and Isufi, E. and Barbarossa, S. and Leus, G.},
  journal={IEEE Trans. Signal Process.},
  volume={66},
  number={13},
  pages={3584--3598},
  year={2018},
  publisher={IEEE}
}

@inproceedings{schaub2018flow,
  title={Flow smoothing and denoising: Graph signal processing in the edge-space},
  author={Schaub, M. T. and Segarra, S.},
  booktitle={GlobalSIP},
  pages={735--739},
  year={2018}}

@article{Panagopoulos_2021_covid, 
title={Transfer Graph Neural Networks for Pandemic Forecasting}, 
volume={35}, 
number={6}, 
journal={AAAI}, 
author={Panagopoulos, G. and Nikolentzos, G. and Vazirgiannis, M.}, 
year={2021}, 
month={May}, 
pages={4838-4845}}

@article{zhao2023sequential,
  title={Sequential {Monte Carlo} Graph Convolutional Network for Dynamic Brain Connectivity},
  author={Zhao, F. and Kuruoglu, E. E.},
  journal={ICASSP},
  year={2024}
}

@ARTICLE{brain_modeling,  author={Huang, W. and Bolton, T. A. W. and Medaglia, J. D. and Bassett, D. S. and Ribeiro, A. and Van De Ville, D.},  journal={Proc. IEEE},   title={A Graph Signal Processing Perspective on Functional Brain Imaging}, pages = {868 - 885},  year={2018},  volume={106},  number={5}}

@INPROCEEDINGS{Krishnan_2023_SVAR,
  author={Krishnan, F. and Money, R. and Beferull-Lozano, B. and Isufi, E.},
  booktitle={ICASSP}, 
  title={Simplicial Vector Autoregressive Model For Streaming Edge Flows}, 
  year={2023},
  volume={},
  number={},
  pages={1-5}}

@misc{transportation,
author = {B. Stabler and H. Bar-Gera and E. Sall},
title = {Transportation Networks for Research},
publisher = {GitHub},
url = {https://github.com/bstabler/TransportationNetworks},
year={2016}
}

@ARTICLE{Barbarossa_2020,
  author={Barbarossa, S. and Sardellitti, S.},
  journal={IEEE Trans. Signal Process.}, 
  title={Topological Signal Processing Over Simplicial Complexes}, 
  year={2020},
  volume={68},
  number={},
  pages={2992-3007}}

@ARTICLE{Yang_2022_Simplicial,
  author={Yang, M. and Isufi, E. and Schaub, M. T. and Leus, G.},
  journal={IEEE Trans. Signal Process.}, 
  title={Simplicial Convolutional Filters}, 
  year={2022},
  volume={70},
  number={},
  pages={4633-4648}}

@ARTICLE{Ortega_2018,  author={Ortega, A. and Frossard, P. and Kovačević, J. and Moura, J. M. F. and Vandergheynst, Pierre},  journal={Proc. IEEE},   title={Graph Signal Processing: Overview, Challenges, and Applications},   
year={2018},  
volume={106},  
number={5},  
pages={808-828}}

@ARTICLE{bib_LMS,  author={D. Lorenzo, P. and Barbarossa, S. and Banelli, P. and Sardellitti, S.},  journal={IEEE Trans. Signal Inf. Process. Netw.},   title={Adaptive Least Mean Squares Estimation of Graph Signals},   year={2016},  volume={2},  number={4}, pages={555 - 568}}

@article{bib_NLMS,
title = {Normalized {LMS} algorithm and data-selective strategies for adaptive graph signal estimation},
journal = {Signal Processing},
volume = {167},
year = {2020},
pages = {107326},
author = {M. J. M. Spelta and W. A. Martins}}

@article{bib_LMP,
title = {Adaptive estimation and sparse sampling for graph signals in alpha-stable noise},
journal = {Digital Signal Processing},
volume = {105},
pages = {102782},
year = {2020},
author = {N. H. Nguyen and K. Doğançay and W. Wang}
}

@ARTICLE{yan_2022_NLMP,
  author={Yan, Y. and Adel, R. and E. E. Kuruoglu},
  journal={J. Signal Process. Syst.}, 
  title={Graph Normalized-{LMP} Algorithm for Signal Estimation Under Impulsive Noise}, 
  year={2022},
  volume={},
  number={},
  pages={}}

@article{yan_2022_sign,
title = {Adaptive sign algorithm for graph signal processing},
journal = {Signal Processing},
volume = {200},
pages = {108662},
year = {2022},
issn = {0165-1684},
author = {Y. Yan and E. E. Kuruoglu and M. A. Altinkaya}
}

@article{Schaub_2020_Random,
author = {Schaub, M. T. and Benson, Austin R. and Horn, P. and Lippner, G. and Jadbabaie, A.},
title = {Random Walks on Simplicial Complexes and the Normalized Hodge 1-Laplacian},
journal = {SIAM Review},
volume = {62},
number = {2},
pages = {353-391},
year = {2020}
}

@inproceedings{kuruoglu1997new,
  title={A new analytic representation for the $\alpha$-stable probability density function},
  author={Kuruoglu, E. E. and Molina, C. and Godsill, S. J. and Fitzgerald, W.J.},
  booktitle={ISBA},
  year={1997}
}
\end{document}